# Analysis of WiMAX Physical Layer Using Spatial Multiplexing

Pavani Sanghoi[#1], Lavish Kansal[*2],

[#1]Student, Department of Electronics and Communication Engineering,
Lovely Professional University, Punjab, India
[1] pavani.sanghoi@gmail.com

[*2]Assistant Professor, Department of Electronics and Communication Engineering,
Lovely Professional University, Punjab, India
[2] lavish.s690@gmail.com

***Abstract:*** Broadband Wireless Access (BWA) has emerged as a promising solution for providing last mile internet access technology to provide high speed internet access to the users in the residential as well as in the small and medium sized enterprise sectors. IEEE 802.16e is one of the most promising and attractive candidate among the emerging technologies for broadband wireless access. The emergence of WiMAX protocol has attracted various interests from almost all the fields of wireless communications. MIMO systems which are created according to the IEEE 802.16-2005 standard (WiMAX) under different fading channels can be implemented to get the benefits of both the MIMO and WiMAX technologies. In this paper analysis of higher level of modulations (i.e. M-PSK and M-QAM for different values of M) with different code rates and on WiMAX-MIMO system is presented for Rayleigh channel by focusing on spatial multiplexing MIMO technique. Signal-to Noise Ratio (SNR) vs Bit Error Rate (BER) analysis has been done.

**Keywords:** BWA, WiMAX, OFDM, ISI, LOS, NLOS, MIMO, SNR, BER, FEC, CC

## I. INTRODUCTION

Worldwide Interoperability for Microwave Access (WiMAX) is an IEEE 802.16 standard based technology responsible for bringing the Broadband Wireless Access (BWA) to the world as an alternative to wired broadband. The IEEE 802.16e air interface standard [1] is basically based on technology namely, orthogonal frequency-division multiplexing (OFDM), that has been regarded as an efficient way to combat the inter-symbol interference (ISI) for its performance over frequency selective channels for the broadband wireless networks. The WiMAX standard 802.1 6e provides fixed, nomadic, portable and mobile wireless broadband connectivity without the need for direct line-of-sight with the base station. WiMAX distinguishes itself from the previous versions of the standard in the sense that this standard adds mobility to the wireless broadband standard.

WiMAX can be classified into Fixed WiMAX [2] and Mobile WiMAX. Fixed WiMAX is based upon Line Of Sight (LOS) condition in the frequency range of 10-66GHz whereas Mobile WiMAX is based upon Non-Line of Sight (NLOS) condition that works in 2-11 GHz frequency range [3]. For 802.16e standard, MAC layer & PHY layer has been defined, but in this paper, emphasis is given only on the PHY layer. PHY layer for mobile WiMAX which is IEEE-802.16e standard [4] has scalable FFT size i.e. 128-2048 point FFT with OFDMA, Range varies from 1.6 to 5 Km at 5Mbps in 5MHz channel BW, supporting 100Km/hr speed.

Multi-Input Multi-Output (MIMO) technology has also been renowned as an important technique for achieving an increase in the overall capacity of wireless communication systems. In this multiple antennas are employed at the transmitter side as well as the receiver side [5]. One can achieve spatial multiplexing gain in MIMO systems realized by transmitting independent information from the individual antennas, and interference reduction. The enormous values of the spatial multiplexing or capacity gain achieved by MIMO Spatial multiplexing technique had a major impact on the introduction of MIMO technology in wireless communication systems.

The paper is organized as follows: Model of WiMAX PHY layer is explained in section II. An overview of the MIMO systems is presented in Section III. MIMO Techniques are provided in section IV. WiMAX-MIMO systems are studied in section V. Results and simulations are shown in section VI. At the end conclusion is given in section VII.

## II. WIMAX MODEL FOR PHYSICAL LAYER

The main task of the physical layer is to process data frames delivered from upper layers to a suitable format for the wireless channel. This task is done by the following processing: channel estimation, FEC (Forward Error Correction) coding, modulation, mapping in OFDMA (Orthogonal Frequency Division Multiple Access) symbols, etc [6]. The block diagram for WiMAX system (802.16e standard) is shown in Figure 1.

**A. Randomization**

It is the first process which is carried out in the WiMAX Physical layer after the data packet is received from the higher layers and each of the burst in Downlink as well as in the Uplink is randomized. It is basically scrambling of data to generate random sequence in order to improve coding performance and data integrity of the input bits [7].

**B. Forward Error Correction (FEC)**

It basically deals with the detection and correction of errors due to path loss and fading that leads to distortion in the signal. There are number of coding systems that are involved in the FEC process like RS codes, convolution codes, Turbo codes, etc. Basically we will be focusing upon the RS as well as the convolution codes.

**1. RS codes**

These are non-binary cyclic codes that add redundancy to the data. This redundancy is basically addition of parity bits into the input bit stream that improves the block errors.

**2. Convolution codes (CC)**

These CC codes introduce redundant bits in the data stream with the use of linear shift registers (m). The information bits are applied as input to shift register and the output encoded bits are obtained with the use of modulo-2 addition of the input information bits. The contents of the shift register in 802.11a physical layer uses Convolution code as the mandatory FEC [8].

These convolutional codes are used to correct the random errors and are easy to implement than RS codes. Coding rate is defined as the ratio of the input bits to the output bits. Higher rates like 2/3 and 3/4, are derived from it by employing "puncturing." Puncturing is a procedure that involves omitting of some of the encoded bits in the transmitter thus reducing the number of transmitted bits and hence increasing the coding rate of the CC code and inserting a dummy "zero" metric into the convolution viterbi decoder on the receive side of WiMAX Physical layer in place of the omitted bits. Code rate of convolution encoder is given as:-

$$\text{Code rate} = k/n$$

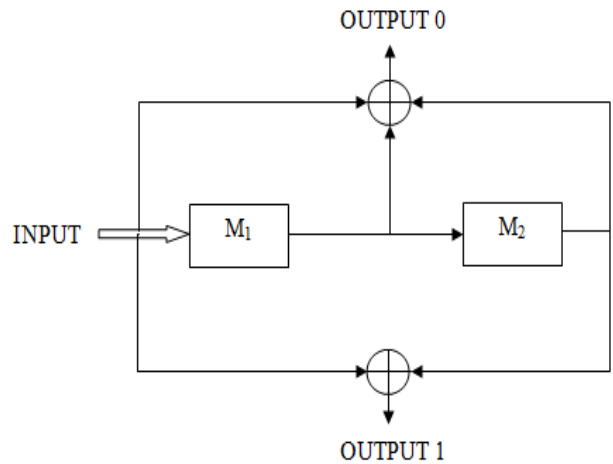

Fig. 1. Convolution Encoder; code rate=1/2, m=2

For Decoding the Viterbi algorithm is used at the receiver side of the PHY layer. To describe a convolution code, one need to characterize the encoding function (m), so that given an input sequence m, one can readily compute the output sequence U.

### C. Interleaving

It aims at distributing transmitted bits in time or frequency domain or both to achieve desirable bit error distribution after the demodulation process. In interleaving, data is mapped onto non-adjacent subcarriers to overcome the effects of multipath distortion and burst errors. Block interleaving mainly operates on one of the block of bits at a time. The number of bits in each block is known as interleaving depth, which defines the delay introduced by interleaving process at the transmitter side. A block interleaver can be described as a matrix to which data is written in column format and data is read in row wise format, or vice versa.

### D. Modulation

This process involves mapping of digital information onto analog form such that it can be transmitted over the channel. A modulator is involved in every digital communication system that performs the task of modulation. Modulation can be done by changing the amplitude, phase, as well as the frequency of a sinusoidal carrier. In this paper we are concerned with the digital modulation techniques. Various digital modulation techniques can be used for data transmission, such as M-PSK and M-QAM, where M is the number of constellation points in the constellation diagram. Inverse process of modulation called demodulation is done at the receiver side to recover the original transmitted digital information.

**E. Pilot Insertion**

Used for channel estimation & synchronization purpose. In this step, pilot carriers are inserted whose magnitude and phase is known to the receiver.

**F. Inverse Fast Fourier Transform (IFFT)**

An Inverse Fast Fourier transform converts the input data stream from frequency domain to time domain representing OFDM Subcarrier as the channel is basically in time domain. IFFT is useful for OFDM system as it generates samples of a waveform with frequency components satisfying the orthogonality condition such that no interference occurs in the subcarriers.

Similarly FFT converts the time domain to frequency domain as basically we have to work in frequency domain [9]. By calculating the outputs simultaneously and taking advantage of the cyclic properties of the multipliers FFT techniques reduce the number of computations to the order of $N \log N$. The FFT is most efficient when $N$ is a power of two.

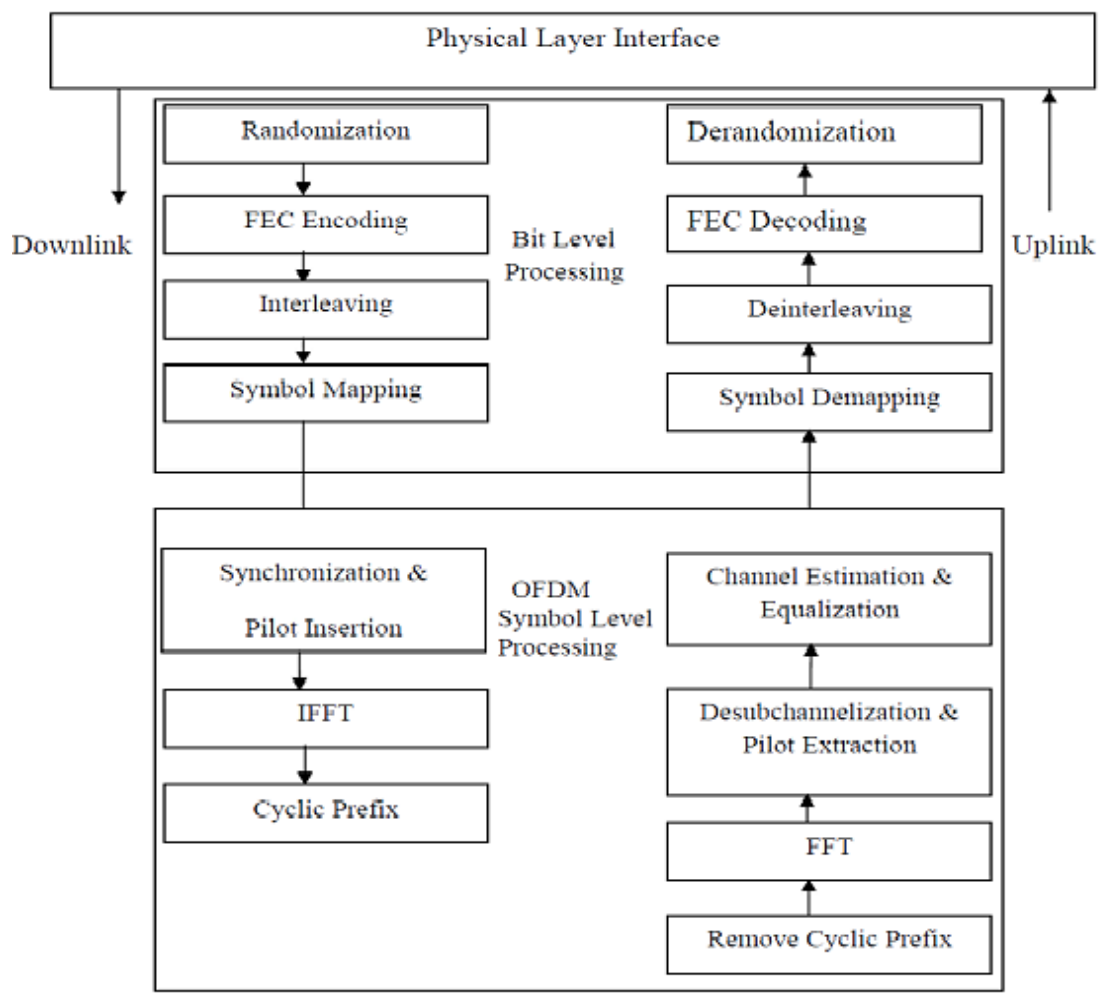

Fig. 2. WiMAX Model for Physical Layer (802.16)

### G. Cyclic Prefix

One way to prevent ISI is basically to create a cyclically extended guard interval in between the data bits, where each of the OFDM symbol is preceded by a periodic extension of the signal itself which is known as the Cyclic Prefix as shown in fig. 3. When the guard interval is longer than the channel impulse response, or the multipath delay, the IS1 can be eliminated.

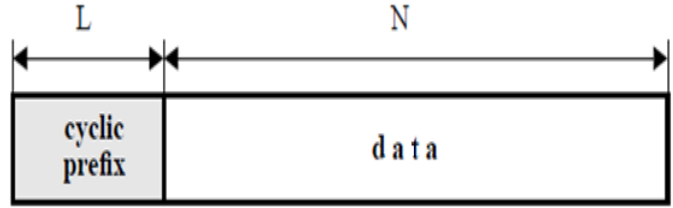

Fig. 3. Cyclic Prefix

### H. Communication Channels

Communication channels are kind of medium of communication between transmitter and receiver. These channels are mainly divided into fast and slow fading channels. A channel is known as fast fading if the impulse response of the channel changes approximately at the symbol rate of the communication system, whereas in slow fading channel, impulse response stays unchanged for several symbols.

Rayleigh channel is used for the analysis purpose in this paper. Constructive and destructive nature of multipath components in flat fading channels can be approximated by Rayleigh distribution if there is no line of sight which means when there is no direct path between transmitter and receiver [10]. The received signal can be simplified to:

$$r(t) = s(t)*h(t) + n(t)$$

where h (t) is the random channel matrix having Rayleigh distribution and n(t) is the Additive White Gaussian noise. The Rayleigh distribution is regarded as the magnitude of the sum of two equal independent orthogonal Gaussian random variables and is useful in LOS condition only.

### I. Receiver Side

The inverse processes take place at the receiver side. Removing the guard interval becomes equivalent to removing the cyclic prefix. Performing a FFT on the received samples after the cyclic prefix is discarded; the periodic convolution is transformed into multiplication, as it was the case for the analog Multi Carrier receiver [11]. Then demodulation, deinterleaving as well as FEC decoding using Viterbi Decoder and at last derandomization.

## III. Multi Input Multi Output (MIMO) Systems

MIMO is an antenna technology for wireless communications in which multiple antennas are used at both the source (transmitter) and the destination (receiver). The antennas at each end of the communications circuit are combined to minimize errors and optimize data speed. Multi-antenna systems can be classified into three main categories. Multiple antennas used at the transmitter side of MIMO systems are mainly used for beamforming purposes to avoid the signal going to undesired directions. Transmitter or the receiver side multiple antennas are used for realizing different (frequency, space) diversity schemes in order to get diversity or capacity gain. The third class includes systems with multiple transmitter and receiver antennas that help in realizing spatial multiplexing which is often referred as MIMO by itself.

In radio communications, MIMO means employing multiple antennas at both the transmitter and receiver side of a specific radio link [12]. In case of spatial multiplexing technique, different data symbols are transmitted through different antennas with the same frequency within the same time interval. Multipath propagation phenomenon is assumed in order to ensure that there is correct operation of spatial multiplexing, since MIMO performs better in terms of channel capacity in a rich multipath scattering environment.

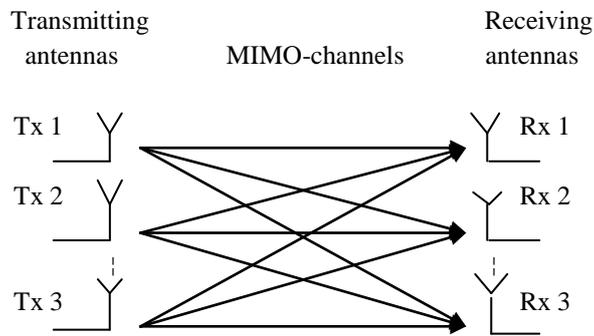

Fig. 4. Block Diagram of MIMO system with M Transmitters and N Receivers

## IV. MIMO Techniques

### A. Spatial Diversity

Diversity is one of the ways to combat the phenomenon of multipath fading. The main idea behind the diversity technique is to provide multiple replicas of the transmitted signal to the receiver. If these multiple replicas fade independently with each other, there is less probability for having all copies of the transmitted signal in deep fade simultaneously [13]. MIMO system takes advantage of the spatial diversity technique by placing independent separate antennas in a dense multipath scattering environment. In spatial diversity, same information is being sent on the independent individual antennas at the transmitter side. These systems can be implemented in a number of ways to obtain a diversity gain to combat with signal fading or a capacity improvement can also be done. Thus, the receiver can easily decode the transmitted signal using these received signals. This technique involves Space Time Block Coding (STBC) and Space Time Trellis Coding (STTC). Diversity techniques can be implemented into different ways in order to improve the bit error rate of the system [14].

### B. Spatial Multiplexing

This form of MIMO is used to provide additional data capacity by utilizing the different paths to carry additional traffic, i.e. increasing the data throughput capability. The spatial multiplexing mitigates the multipath propagation phenomenon problem that is experienced by most of the microwave transmissions. MIMO techniques permit multiple streams that help in improving the signal-to-noise ratio and also the reliability that significantly improves over other versions of the standard. Spatial multiplexing includes transmitting different information onto different independent individual antennas at the transmitter side of the MIMO system and thus helps in attaining the capacity gain. This technique includes V-BLAST technology that is used to improve the spectral efficiency of the system [15].

MIMO systems utilize spatial multiplexing under rich scattering environment; independent data streams are simultaneously transmitted over different antennas to increase the effective data rate. MIMO spatial multiplexing [16] requires at least 2 transmitters and 2 receivers, and the receivers must be in the same place that means they should be in the same device. Because the transmitting antennas are not required to be in the same device and also two mobiles can be used together in the uplink.

**C. Beamforming**

Beamforming enables performance gains with multiple antennas at the BS and even a single antenna at the MS so as to direct the beam in a particular direction such that the signal going in the desired direction is increased and the signal going to the other directions is decreased. Such performance gains are derived from the array gain obtained from the phased array antennas used in the beamforming plus the diversity gain, which can be as much as 10 dB for a system having four antennas at the base station and a single antenna at the mobile station. Beamforming is a signal processing technique used to control the directionality of the transmission and reception of radio signals. The most effective and efficient type of beamforming is dynamic digital beamforming [17]. This uses an advanced, on-chip digital signal processing (DSP) algorithm in order to gain complete control over all the Wi-Fi signals.

# V. WiMAX-MIMO Systems

MIMO systems created according to the IEEE 802.16-2005 standard (WiMAX) under different fading channels can be implemented to get the benefits of both the MIMO and WiMAX technologies. Main aim of combining both WiMAX and Spatial multiplexing MIMO technique is to achieve higher data rates by lowering the BER and improving the SNR of the whole system. The proposed block diagram of WiMAX-MIMO systems is given in Figure 5.

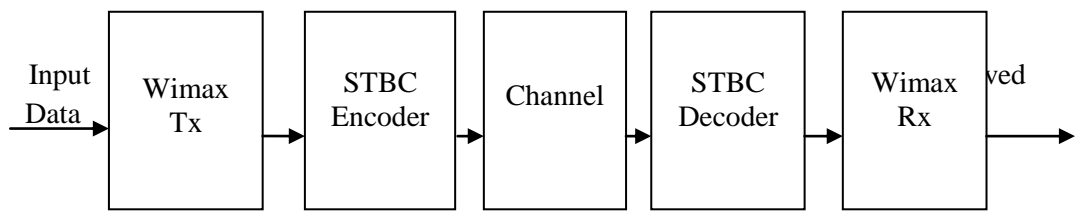

Fig. 5. WiMAX-MIMO System

The use of WiMAX technology with the MIMO technology provides an attractive solution for future broadband wireless systems that require reliable, efficient and high-rate data transmission. Employing MIMO systems in WiMAX [18] yields better BER performance compared to simple WiMAX protocol. Spatial multiplexing technique of MIMO systems provides spatial multiplexing gain that has a major impact on the introduction of MIMO technology in wireless systems thus improving the capacity of the system. Combining of both the systems involves employing STBC encoder and decoder at the transmitter and receiver side of WiMAX Physical Layer respectively.

This paper analyze the WiMAX protocol as well as the spatial multiplexing technique of MIMO systems in order to achieve higher data rates by lowering the Bit Error Rate and improving the SNR value of the system to achieve better performance and results. Spatial multiplexing (SM) is a recently developed transmission technique that uses multiple antennas and helps in achieving the capacity gain.

## VI. Results and Simulations

In this paper behavior of the WiMAX-MIMO system under different modulations with different convolutional code rates is studied and the effects of increasing the order of the modulation on the BER performance of the system are presented. Results are shown in the form of SNR vs BER plot for different modulations that shows an improvement in the SNR value when we employ spatial multiplexing technique of MIMO system with the WiMAX protocol. In the case of spatial multiplexing, capacity improvement can also be seen as we are sending different data over the independent individual antennas. The analysis has been done for Rayleigh channel.

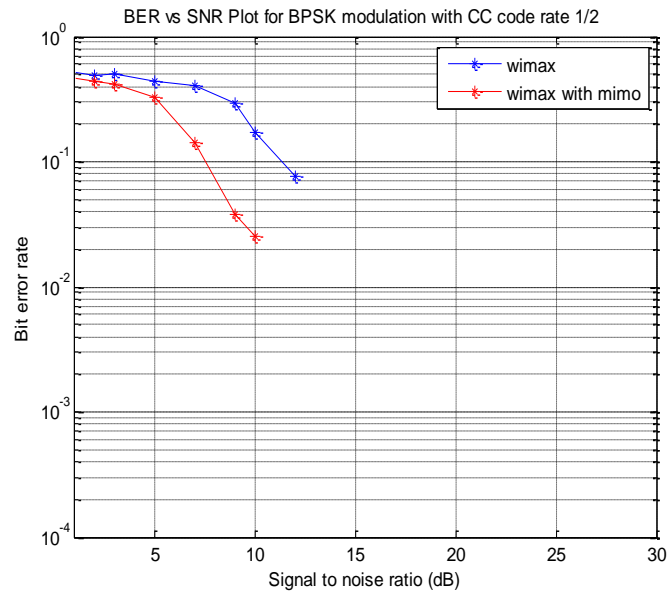

(a) BPSK with CC code rate 1/2

In this graph we are able to get 2dB improvement in SNR and capacity improvement is also shown when we employ SM technique of MIMO in WiMAX in the presence of Rayleigh channel.

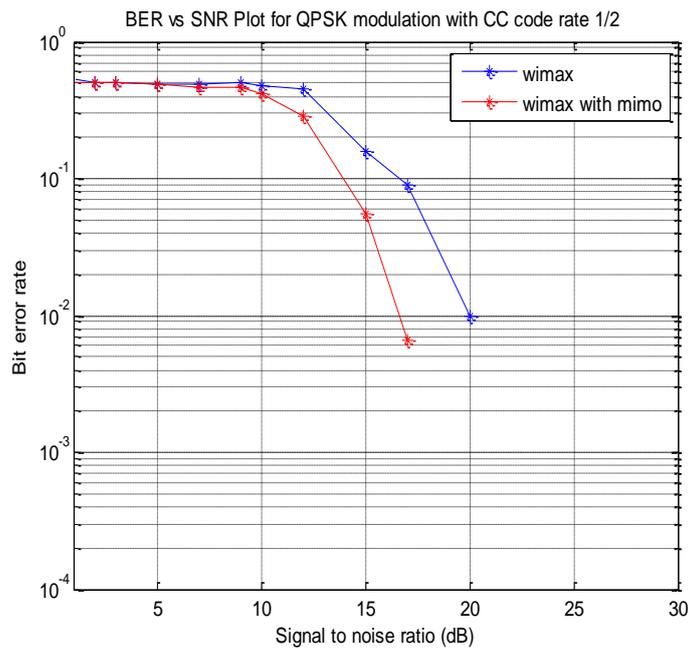

(b) QPSK with CC code rate 1/2

QPSK modulation with CC code rate 1/2 provides 2dB improvement in the SNR value when we combine WiMAX with MIMO SM technique as well as capacity gain can also be seen. Rayleigh channel is used for the analysis.

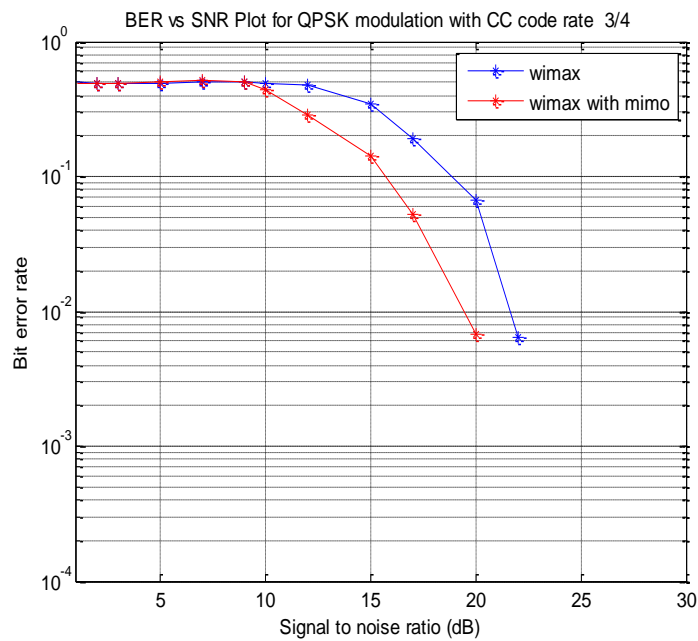

(c) QPSK with CC code rate 3/4

Capacity as well as SNR improvement of 2dB can be seen when we use QPSK modulation with CC code rate of 3/4 in WiMAX-MIMO system compared to simple WiMAX using Rayleigh channel.

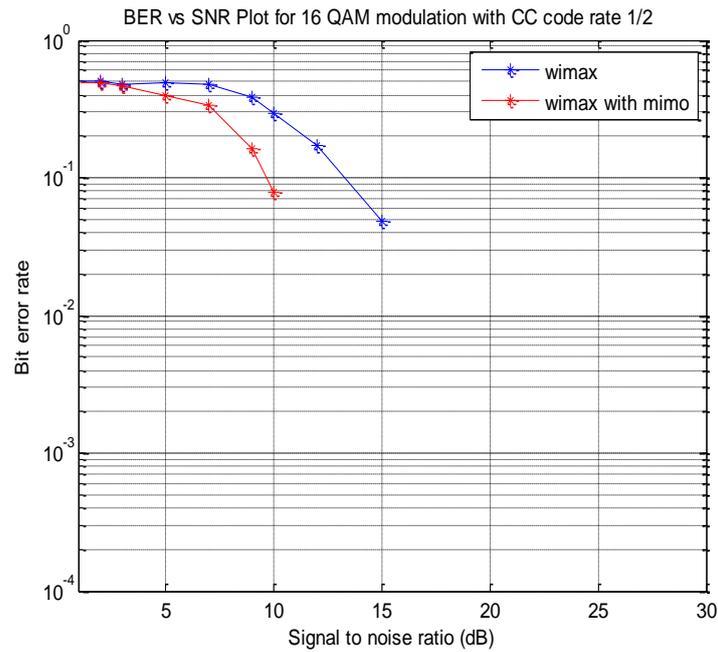

(d) 16QAM with CC code rate 1/2

As seen from the graph, SNR improvement of 5dB and capacity improvement is there using WiMAX with SM technique of MIMO technology in the presence of Rayleigh channel.

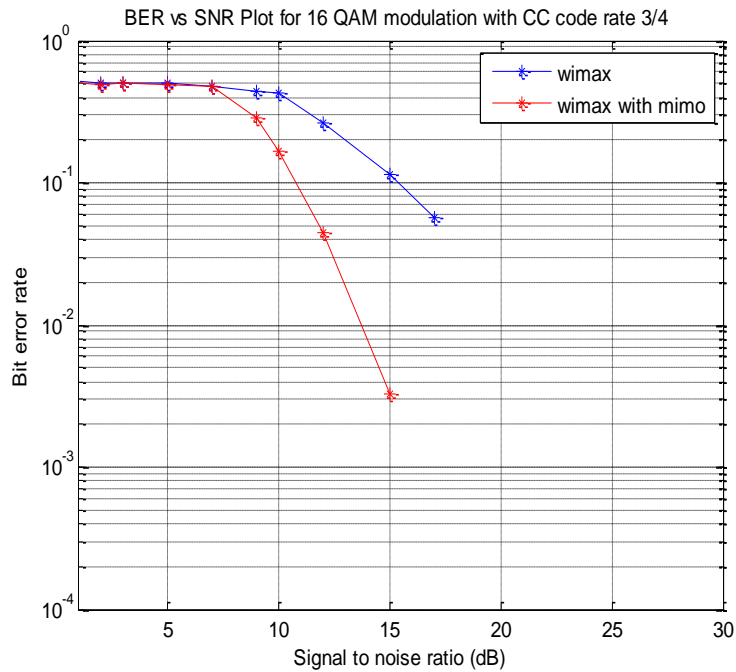

(e) 16 QAM with CC code rate 3/4

Again as we can see there is an improvement of 2dB in SNR using Rayleigh channel with combined effect of WiMAX-MIMO technology.

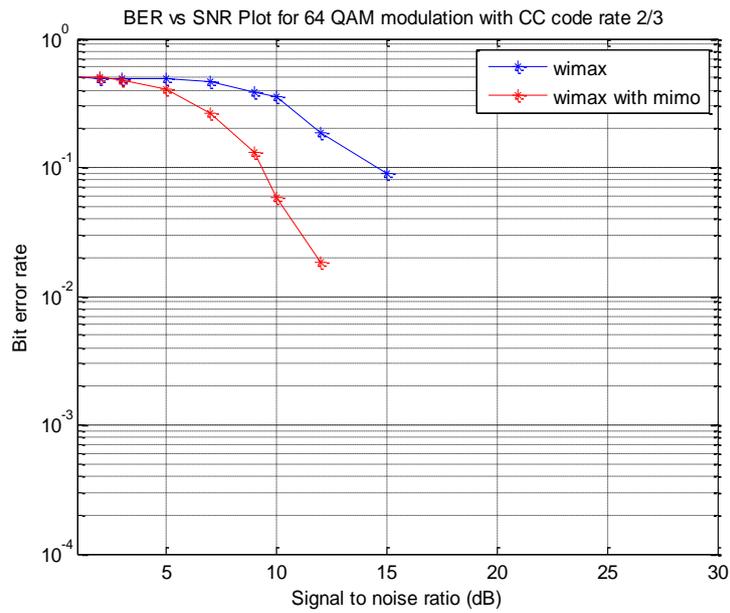

(f) 64 QAM with CC code rate 2/3

SNR improvement of 3dB in the case of 64QAM with CC code rate 2/3 can be seen and capacity improvement can also be seen in when we use WiMAX-MIMO in the presence of Rayleigh channel.

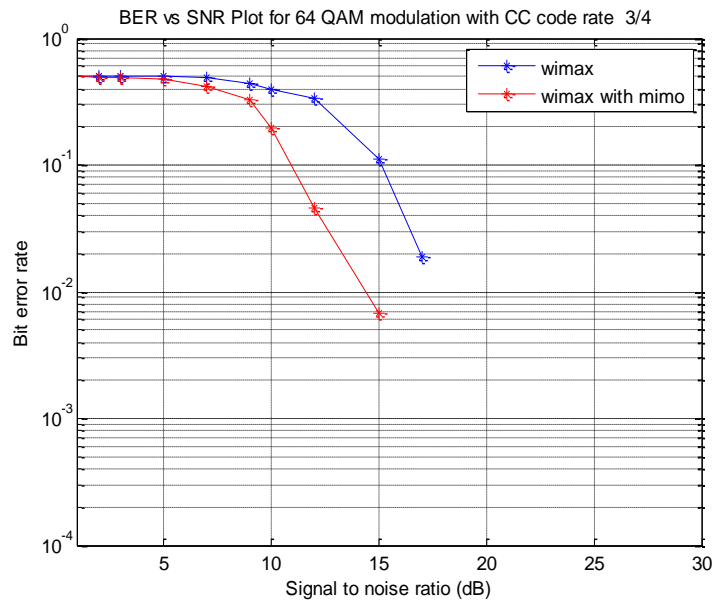

(g) 64 QAM with CC code rate 3/4

This graph shows an SNR improvement of 2dB as well as capacity gain using WiMAX-MIMO system in the presence of Rayleigh channel.

Fig. 6. BER vs SNR plots for Rayleigh channel
a) BPSK code rate 1/2 b) QPSK code rate 1/2 c) QPSK code rate 3/4 d) 16 QAM code rate 1/2
e) 16 QAM code rate 3/4 (f) 64 QAM code rate 2/3 (g) 64 QAM code rate 3/4

The performance in the form of BER vs SNR plots for different modulations over Rayleigh channel for WiMAX-MIMO system with different CC code rates have been presented in Figure 6 (a)–(g). Each graph shows an improvement in SNR using spatial multiplexing technique of MIMO system which is given in the following table.

Table 1: SNR improvement in Rayleigh channel by using Spatial Multiplexing in WiMAX

| MODULATION | SNR Improvement using Rayleigh channel (dB) |
|---|---|
| BPSK code rate 1/2 | 2dB |
| QPSK code rate 1/2 | 3dB |
| QPSK code rate 3/4 | 2dB |
| 16 QAM code rate 1/2 | 5dB |
| 16 QAM code rate 3/4 | 2dB |
| 64QAM code rate 2/3 | 3dB |
| 64QAM code rate 3/4 | 2dB |

## VII. CONCLUSIONS

In this paper effect of employing spatial multiplexing technique of MIMO system in WiMAX 802.16e PHY layer has been simulated through Matlab2009a. This technique of MIMO systems provides spatial multiplexing gain that has a major impact on the introduction of MIMO technology in wireless systems. Rayleigh channel has been taken into account for the analysis purpose. Simulations are based upon using different modulations with different convolutional code rates and show that there is improvement in the SNR value as well as capacity improvement can also be seen by employing spatial multiplexing technique of MIMO system in WiMAX protocol.

Results are presented in the form of BER vs SNR value and show that BER reduces when we employ MIMO system in WiMAX in comparison to simple WiMAX. This shows that employing MIMO system in WiMAX improves the overall performance of the system and provides capacity gain. Main aim is to reduce the BER of the system for lower value of SNR hence providing higher data rates for the transmission purpose such that originality of the input signal is retained.

**Authors**

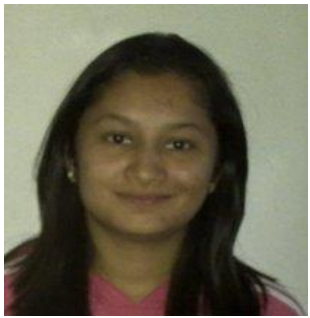

**Pavani Sanghoi** was born in Jammu (Jammu and Kashmir). She has done her B.tech degree in Electronics and Communication Engineering from Lovely Professional University, Jalandhar. She is currently pursuing her M.tech degree (last semester) from Lovely Professional University, Jalandhar. Her research interest includes Digital Communication and Wireless Communication.

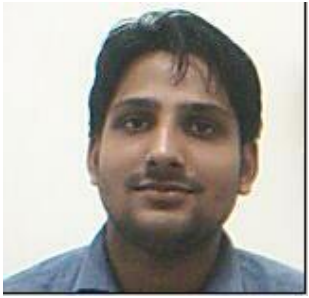# 
**Lavish Kansal** was born in Bathinda (Punjab). He received his BTech. Degree in Electronics and Communication Engineering from PTU, Jalandhar. He received his degree of ME from Thapar University, Patiala. He is currently appointed as Assistant Professor in Lovely Professional University, Jalandhar, Punjab. His research interest includes Digital Signal Processing, Digital Communication and Wireless Communication.